\documentclass[a4paper,fleqn,usenatbib]{mnras}

\usepackage{newtxtext,newtxmath}

\usepackage[T1]{fontenc}
\usepackage{ae,aecompl}

\usepackage{amsmath}	
\usepackage{amssymb}	
\usepackage{graphicx,color,afterpage}
\usepackage{times,subcaption,rotating,tikz}
\usepackage{float}

\def\hii{\mbox{H\sc{ii}}}

\def\msun{M$_{\odot}$}

\def\arcsec{$^{\prime \prime}$}
\definecolor{Mygrey}{gray}{0.75}

\newcommand{\gtsimeq}{\raisebox{-0.6ex}{$\,\stackrel{\raisebox{-.2ex}{$\textstyle >$}}{\sim}\,$}}

\newcommand{\deltaoh}{$\Delta\left({\rm O/H}\right)$}

\mathchardef\mhyphen="2D

\usepackage[compact]{titlesec}
\titlespacing{\section}{0pt}{*2}{*1}

\title[Gas accretion in ETGs]{Gas accretion as fuel for residual star formation in Galaxy Zoo elliptical galaxies} 
\author[Timothy A. Davis]{\parbox{\textwidth}{Timothy A. Davis$^{1}$\thanks{E-mail: \texttt{DavisT@cardiff.ac.uk}} and Lisa M. Young$^{2,3}$}
\vspace{0.3cm}
\\
\parbox{\textwidth}{$^{1}$School of Physics \&\ Astronomy, Cardiff University, Queens Buildings, The Parade, Cardiff, CF24 3AA, UK\\
$^{2}$Physics Department, New Mexico Tech, 801 Leroy Place, Socorro, NM 87801, USA\\
$^{3}$Adjunct Astronomer, National Radio Astronomy Observatory, Socorro, NM 87801, USA}}
\begin{document}
\date{Accepted 2019 September 3. Received 2019 August 20; in original form 2019 July 11.}

\pagerange{\pageref{firstpage}--\pageref{lastpage}} \pubyear{2019}

\maketitle

\label{firstpage}

\begin{abstract}
In this letter we construct a large sample of early-type galaxies with measured gas-phase metallicities from the Sloan Digital Sky Survey and Galaxy Zoo in order to investigate the origin of the gas that fuels their residual star formation. 
We use this sample to show that star forming elliptical galaxies have a substantially different gas-phase metallicity distribution from spiral galaxies, with $\approx$7.4\% having a very low gas-phase metallicity for their mass. 
These systems typically have fewer metals in the gas phase than they do in their stellar photospheres, which strongly suggests that the material fuelling their recent star formation was accreted from an external source. 
We use a chemical evolution model to show that the enrichment timescale for low-metallicity gas is very short, and thus that cosmological accretion and minor mergers are likely to supply the gas in \gtsimeq37\% of star-forming ETGs, in good agreement with estimates derived from other independent techniques. 
\end{abstract}

\begin{keywords}
galaxies: elliptical and lenticular, cD -- galaxies: spiral --  galaxies: abundances -- galaxies: evolution -- galaxies: star formation 
\end{keywords}

\section{Introduction}
\label{intro}

Early-type (lenticular and elliptical) galaxies (ETGs) in the local universe are typically passive systems \citep[e.g.][]{2005ApJ...621..673T}, lying well below the star formation main sequence \citep[e.g.][]{2004MNRAS.351.1151B,2015ApJ...808L..49G}, and thus not forming (cosmologically) significant amounts of new stars. It is hard to keep these systems entirely passive for long, however, as accretion of gas from the environment \citep[e.g.][]{2011MNRAS.414.2458V, 2011MNRAS.417.2982F} and the accumulation of stellar mass loss material is inevitable \citep[e.g.][]{1995A&A...298..784G,2001A&A...376...85J,2002ApJ...571..272A}. This makes these systems ideal cosmic laboratories, from which we can study these processes, and how they may be disrupted to keep galaxies passive (e.g. by stellar or AGN feedback; \citealt{2018ApJ...866...70L}).

Around a third of local ETGs have ongoing centrally concentrated (low efficiency) star formation (e.g. \citealt{2007ApJS..173..619K,2009ApJ...695....1T,2011MNRAS.410.1197C,2011MNRAS.415...61S,2013MNRAS.432.1914M,2013MNRAS.432.1796A,2013MNRAS.429..534D,2014MNRAS.444.3427D,2017MNRAS.464.1029N}). 
Understanding the origin of the material fueling this residual star formation is crucial if we wish to construct a complete picture of the evolution of ETGs, and the processes which fuel them. 

Both \textit{internal} and \textit{external} avenues exist for quenched ETGs to obtain gas and resume forming stars. Various studies have set limits on the relative importance of these mechanisms (e.g.\\ \citealt{2006MNRAS.366.1151S,2010MNRAS.404.1775T,2011MNRAS.417..882D,2015MNRAS.449.3503D,2019MNRAS.486.1404D,2014MNRAS.443.1002L,2015MNRAS.448.1271L}; \citealt{2016MNRAS.457..272D,2017MNRAS.466.2570B,2019MNRAS.483..458B,2019MNRAS.484..562G}), but no definitive conclusions have been reached. 

Internal gas replenishment can occur as evolved stars shed enriched material which joins the hot halo of the galaxy. In massive ETGs stellar mass loss rates of more than a solar mass a year are expected \citep{2001A&A...376...85J} which, if it can cool, can easily fuel the observed residual star formation \citep[e.g][]{2018ApJ...853..177P}. However, if this is the dominant source of this material it is then not clear why \textit{all} ETGs do not have low level star formation. The key signature of this gas fuelling mechanism is that the gas will be metal rich and share the angular momentum of the stars 
(as it cools from the inner halo, which has been enriched and torqued to be co-rotating by the stars in the galaxy). If the object was star forming in the relatively recent past then gas pushed into the halo by winds could also fall back, and would likely have similar properties.

External mechanisms exist that bring in gas, and also re-ignite star formation. The most obvious of these is mergers with gas rich galaxies, or accretion of material from the intergalactic medium (IGM). In both of these cases the gas accreted is likely to have a low metallicity, as massive galaxies are much more likely to merge with low-mass systems, and the IGM is metal poor \citep[e.g.][]{2008ApJ...679..194D}. The angular momentum of this gas will initially be uncorrelated with the angular momentum of the stars in the galaxy. 

Many of the studies addressing the origin of the gas in ETGs have concentrated on constraining the angular momentum distribution of the gas (\citealt{2006MNRAS.366.1151S}, \citealt{2011MNRAS.417..882D}, \citealt{2017MNRAS.466.2570B}, \citealt{2019MNRAS.483..458B}), and modelling this in concert with constraints on the large scale environment which can act to suppress external gas accretion (\citealt{2016MNRAS.457..272D,2019MNRAS.486.1404D}; see also \citealt{2010MNRAS.404.1775T}). Less attention has been paid to the metallicity distribution of the gas in ETGs.  This is primarily because gas-phase metallicities derived from optical line emission typically require the ionisation to be dominated by star formation. Given the low star-formation rates in ETGs their ionised gas is often dominated by ionisation from old stellar populations or AGN \citep[e.g.][]{2010MNRAS.402.2187S,2016MNRAS.461.3111B}, which severely limits the available samples. \cite{2019MNRAS.484..562G} study a small sample of ETGs, and find that the metallicity of the gas in their three objects is consistent with a stellar mass loss origin. 

Dust-to-gas ratios can also be used as a proxy for gas-phase metallicity. This technique has identified low dust-to-gas ratios in some ETGs \citep[e.g.][]{2012ApJ...748..123S,2019arXiv190602712L}, and has also been used to reveal the merger origin of the gas in a small number of ETGs with optical signatures of recent minor mergers \citep{2015MNRAS.449.3503D,2018MNRAS.476..122V}, but larger scale studies addressing this issue are missing.

In this letter we report on a targeted study, combining catalogues of gas-phase and stellar metallicities derived from SDSS data \citep{2009ApJS..182..543A} with citizen scientist morphological classifications of each object from Galaxy Zoo \citep{2008MNRAS.389.1179L,2011MNRAS.410..166L}. This allows us to select the largest sample of star-forming elliptical galaxies with measured metallicities to date, which we use to probe the origin of their gas. In Section \ref{data} we discuss these datasets, and their combination. In Section \ref{results} we present our results, before discussing and concluding in Section \ref{discuss}.

 \begin{figure}
 \begin{center}
\includegraphics[width=0.45\textwidth,angle=0,clip,trim=0cm 0cm 0cm 0.0cm]{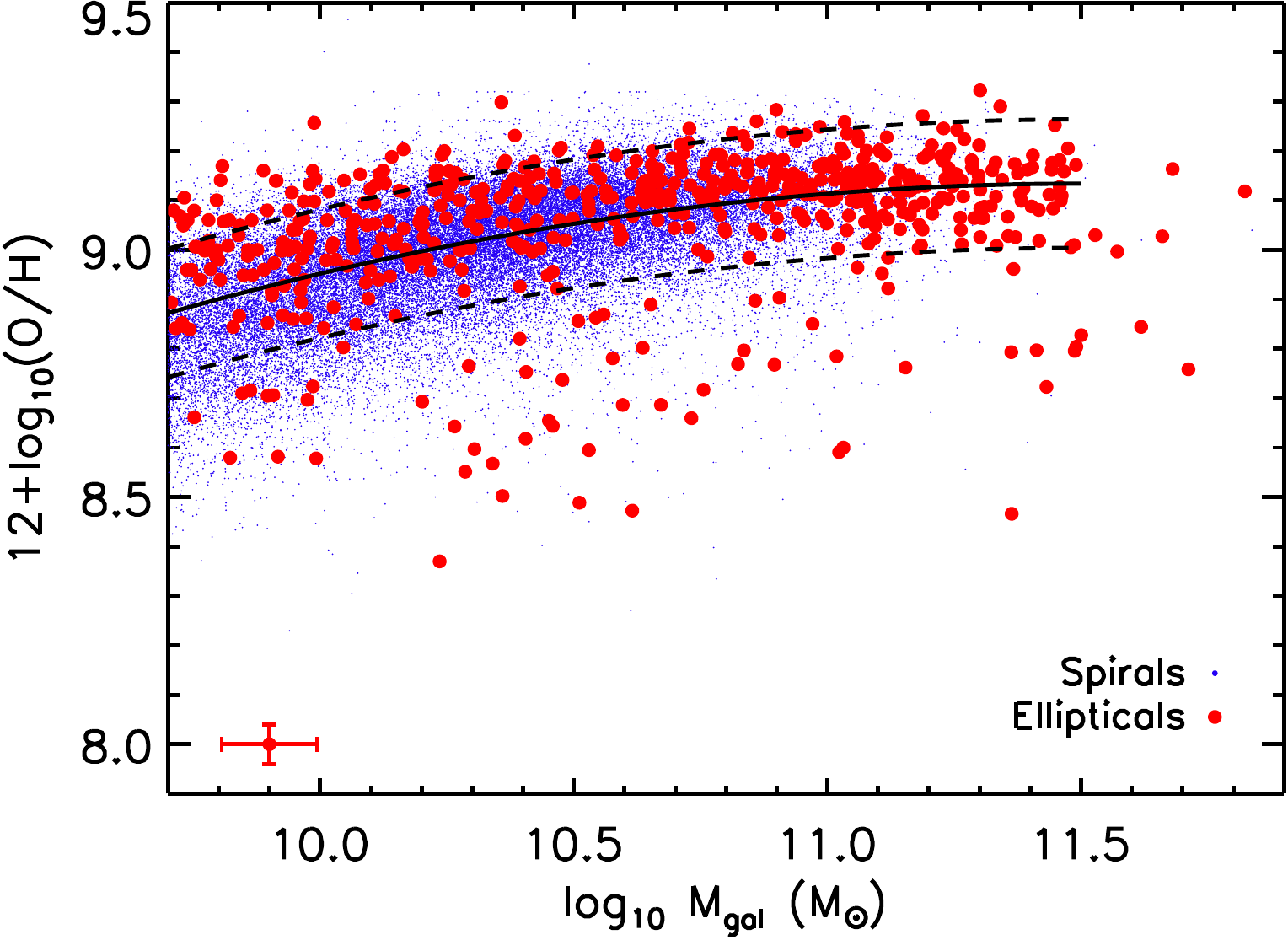}
\caption{Gas-phase metallicity plotted vs stellar mass for Galaxy Zoo classified spiral (blue dots) and elliptical galaxies (red points). The median error for the ETG points is shown in the bottom left corner. The mass -- metallicity relation of {\protect \cite{2004ApJ...613..898T}} is shown as a solid black line, with its 1$\sigma$ scatter shown as dashed lines. A significant population of low gas-phase metallicity ellipticals are found well below the mass -- metallicity relation. }
\label{mass_metal}
 \end{center}
  \vspace{-0.6cm}
 \end{figure}
 
\section{Data Sources}
\label{data}

In order to understand the origin of the gas in ETGs we make used of the Sloan Digital Sky Survey Data Release 7 \citep{2009ApJS..182..543A}, which contains fibre spectra for 929,555 galaxies, obtained using a 3\arcsec\ fibre positioned over the galaxy centre. 
Galaxy masses, gas-phase metallicities and star formation rates for each object in this survey are taken from the Max Planck for Astrophysics--John Hopkins University (MPA-JHU) catalogues \citep{2003MNRAS.341...33K,2004MNRAS.351.1151B,2004ApJ...613..898T}.  These are derived from absorption and emission lines in SDSS DR7 spectra, and have been widely used in the literature. We here select only objects with stellar masses above $5\times10^9$\msun, to avoid including significant numbers of dwarf galaxies.

 \begin{figure} \begin{center}\includegraphics[width=0.45\textwidth,angle=0,clip,trim=0cm 0cm 0cm 0.0cm]{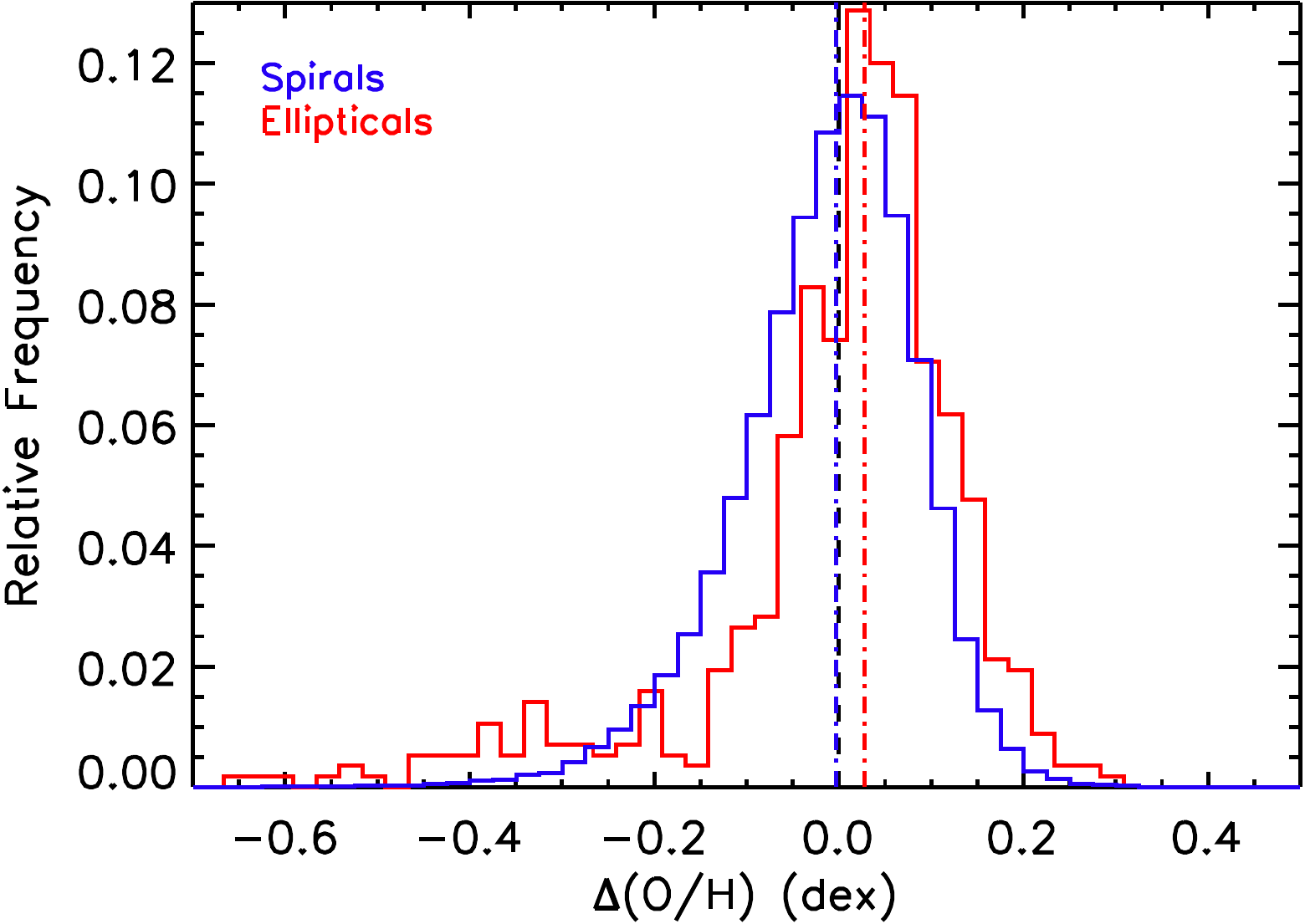}
\caption{Histogram of the residuals around the mass -- metallicity relation of {\protect \cite{2004ApJ...613..898T}} for Galaxy Zoo classified spiral (blue) and elliptical galaxies (red). The black dashed line shows an offset of zero as a guide to the eye, while the blue and red dash-dot lines show the median of the spiral and elliptical galaxy populations, respectively. The average star forming elliptical galaxy is slightly more metal rich than the average star forming spiral galaxy (at fixed mass), but a tail of very low metallicity galaxies also exists, which is significantly more prominent for ellipticals than spirals.}\label{mass_metal_hist}\end{center}\end{figure}

In order to measure gas-phase metallicities ionised gas emission has to be detected within the SDSS fibre, and the gas has to have its ionisation dominated by star formation. 
Fulfilling this condition depends on the density of OB stars in the fibre, and the relative importance of their UV light compared to other sources of radiation. As discussed in \cite{2004ApJ...613..898T} this imposes a somewhat complex selection function which does not depend linearly on any one variable. The vast majority of AGN are removed, as are objects which form stars but in which old stellar populations dominate the ionisation of the gas. 

Metallicities derived from strong emission lines suffer from various uncertainties and biases \citep[e.g.][]{2008ApJ...681.1183K}. Many of these uncertainties are systematic changes between calibrations. In this work we primarily intend to inter-compare different galaxy populations. As such although any given metallicity value quoted may depend on the indicator used, relative differences are likely to be more robust. \hii\ region ageing \citep[e.g.][]{2019arXiv190604520S}, however, primarily affects the metallicity determination in low SFR objects. We tested if this effect is important by excluding objects with H$\alpha$ equivalent widths $<5$\AA, where the effect is strongest, and confirmed that this would not change our results.

In addition to gas-phase metallicities, we collect stellar metallicities for each object from the FIREFLY (Fitting IteRativEly For Likelihood analYsis) database \citep{2017MNRAS.472.4297W}. FIREFLY provides light- and mass-weighted stellar population age and metallicities derived from the SDSS data using stellar population models. We here utilise light-weighted metallicities derived using a \cite{1955ApJ...121..161S} IMF and the MILES stellar models as described in \cite{2011MNRAS.418.2785M}, although we note that our conclusions do not depend on these choices.

\subsection{Morphologies}
To determine the morphology of each galaxy in our sample we use the classifications made by citizen scientists as part of the first data release of the 
Galaxy Zoo (GZ) project \citep{2008MNRAS.389.1179L,2011MNRAS.410..166L}. 

Volunteers classified images of Sloan Digital Sky Survey galaxies as belonging to one of six categories - elliptical, clockwise spiral, anticlockwise spiral, edge-on, star/don't know, or merger.  These classifications include all galaxies which have spectra included in SDSS Data Release 7 and have been extensively used in the literature for a variety of applications. We note that lenticulars, which are typically included in ETG classifications, were not specifically classified by users in GZ data release one. As such the `spiral' and `elliptical' classes likely are contaminated by lenticular galaxies which are preferentially edge- and face-on, respectively.  
For an object to be classified as an elliptical at least 80\% of Galaxy Zoo users must have placed it in this class. A statistical de-biasing correction is included (as discussed in \citealt{2009MNRAS.393.1324B}).

Matching these catalogues leaves us with a total sample of 662,888 objects, of which 61,911 are classified as elliptical galaxies.
Of these 61,911 elliptical galaxies 567 (0.92\%) have stellar masses above $5\times10^9$\,\msun, and measurable gas-phase metallicities (i.e. the ratios of their ionised gas lines are consistent with ionisation from star-formation, based on the criteria of \citealt{2003MNRAS.346.1055K}). 
 These are the objects we will study further in this work. 

{ This sample contains ETGs with stellar masses from $5\times10^9$\,--\,$7\times10^{12}$\,\msun, with redshifts from 0.014 -- 0.30 (although we note that 65\% of our sample is at $z<$\,0.1, likely because of the difficulty in classifying high-redshift objects in SDSS imaging). The lowest SFR probed in our sample is 0.03 \msun\ yr$^{-1}$, and the median is 2 \msun\ yr$^{-1}$. These objects are found throughout the $u$-$r$ colour magnitude diagram, however 3/4 of the sample are bluer than expected for red-sequence objects at this redshift (with $u-r$ colours $<$\,2.4), and thus our sample likely has significant overlap with the smaller sample of 204 blue ETGs selected from Galaxy Zoo by \cite{2009MNRAS.396..818S}.}

\section{Results}
\label{results}

\subsection{Metallicity distribution}

Figure \ref{mass_metal} shows the mass -- metallicity relation for the galaxies classified as spirals and ellipticals in our MPA-JHU + Galaxy Zoo sample (blue and red points, respectively). The median error for the ETG points is shown in the bottom left corner. Also shown as a black line is the mass--metallicity relation of \cite{2004ApJ...613..898T}, and its 1$\sigma$ scatter (black dashed lines). 
The GZ classified ellipticals are present throughout the mass -- metallicity relation, but a sizeable fraction of the ellipticals have significantly lower metallicities.

In order to quantify this further Figure \ref{mass_metal_hist} shows a histogram of \deltaoh\ (the distribution of residuals around the mass--metallicity relation of \citealt{2004ApJ...613..898T}) for our two samples. Spiral galaxies have a reasonably regular distribution of residuals, which peaks around zero and is slightly skewed towards low values. On the other hand, ellipticals have a distribution which peaks at high metallicities, but with a long tail towards lower values. Beyond $\approx$0.3 dex below the mass--metallicity relation the ellipticals are over represented in our dataset. Indeed, 7.4\% (42/567) of our ETGs have metallicities at least 2 standard deviations (0.26 dex) below the \cite{2004ApJ...613..898T} relation, while only 1.7\% (776/45508) of spirals do. We estimate by bootstrapping (with replacement) that the probability of obtaining at least this number of extreme ETG outliers by drawing a sample of 567 objects from the parent distribution of spiral galaxies is $<1\times10^{-6}$. We thus conclude that this excess of low metallicity ETGs is strongly significant. 

{These 42 objects with low metallicity are found at all stellar masses probed in our study, and repeating the statistical analysis described above we are unable to reject the null hypothesis that their stellar masses are simply drawn at random from the parent population. The presence of such objects at all stellar masses is consistent with the conclusions of the majority of studies which select regenerated ETGs by observing gas tracers (see e.g. Section \ref{intro}), but inconsistent with studies which selected regenerating galaxies via proxies of their stellar population age \citep[e.g.][]{2009MNRAS.396..818S,2010MNRAS.404.1775T}. These studies found that blue/young ETGs are predominately lower mass objects. This difference likely arises because of the different systematics inherent in their selection criteria. For an object at fixed distance detection of gas tracers depends only on the mass of regenerating material, while stellar population/colour analyses are sensitive to the fraction of stellar mass formed in the burst compared to the mass of the old stellar component. A star formation burst of a fixed size will more noticeably affect the stellar populations of low-mass ETGs, while even UV colours can fail to pick out evidence of the same amount of star formation buried in the core of a very massive ETG \citep[see e.g.][]{2014MNRAS.444.3408Y}. } 

 In our sample ellipticals also seem over represented at the highest residual metallicities, which may reflect the previously reported correlations between SFR and metallicity (e.g. \citealt{2010MNRAS.408.2115M}).

\subsection{Comparison with stellar metallicities}

In Figure \ref{mass_metal} we identified a population of metal poor ellipticals which lie well below the mass--metallicity relation, and thus are good candidates for having accreted gas. However it is possible that these objects simply have metal poor stellar populations, and accumulated metal poor gas from the the mass loss of these stars.

\begin{figure}
 \begin{center}
\includegraphics[width=0.43\textwidth,angle=0,clip,trim=0cm 0cm 0cm 0.0cm]{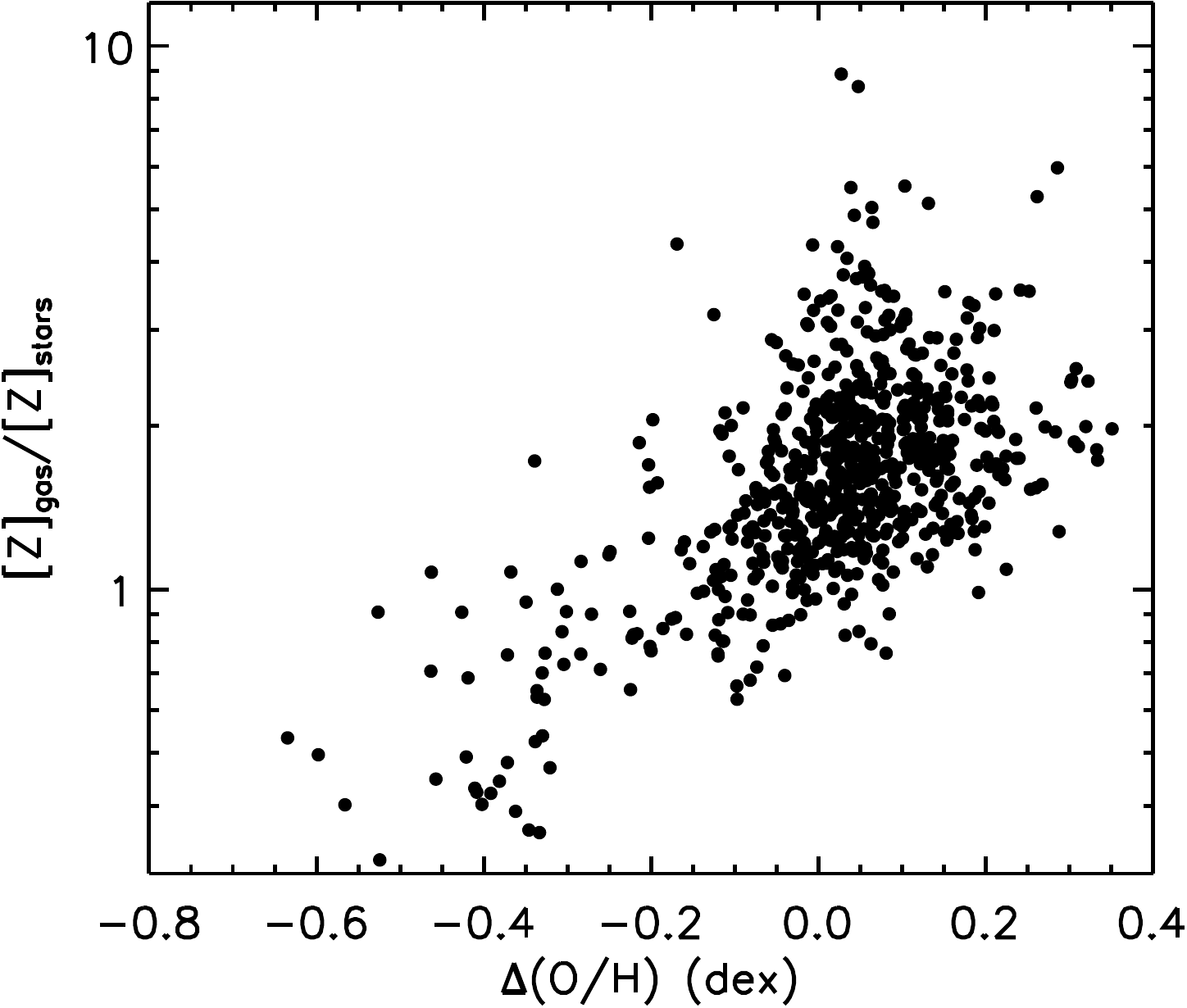}
\caption{Residuals around the mass -- metallicity relation of {\protect \cite{2004ApJ...613..898T}} for Galaxy Zoo classified elliptical galaxies, plotted against the ratio between the gas-phase and stellar metallicities for each object (where each of these metallicities is expressed in solar units, with an assumed oxygen metallicity of the sun of 12+log(O/H)=8.9; \citealt{1989GeCoA..53..197A}). The elliptical galaxies which lie well below the mass -- metallicity relation typical have lower gas-phase than stellar metallicity, a clear signature that this material is not provided by stellar mass loss. }
\label{mass_metal_wstellar}
 \end{center}
  \vspace{-0.6cm}
 \end{figure}

In Figure \ref{mass_metal_wstellar} we plot the the ratio of the gas-phase metallicity to the stellar metallicity for our sample ETGs, plotted as a function of \deltaoh\ (the excess or deficit of metallicity in each galaxy away from that predicted from the mass--metallicity relation of  \citealt{2004ApJ...613..898T}). We note that systematics related to the different methods of measuring the stellar and gas-phase metallicity may scale the ordinate and abscissa values, however they should not change the correlation between the variables.

Stellar mass loss material would be expected to have the same, or (most likely) higher metallicity than estimated for the stars in the galaxy (as the gas is likely to have been enriched since the stars which dominate the stellar metallicity measurement were formed). On the other hand, accreted gas (as it is statistically more likely to come from a lower mass object) can be of lower metallicity than the stellar components of the ETG. Figure \ref{mass_metal_wstellar} shows a clear correlation, with a {Spearmans rank correlation coefficient of 0.45 (which deviates from zero with a significance of $>$10$\sigma$)}.  The majority of galaxies which have \deltaoh$>-0.26$ dex (i.e. they lie within the scatter around the mass--metallicity relation) have gas which is enriched relative to the stars, with a median gas-phase metallicity enhancement of a factor 1.68. While this exact value should be treated with caution, the objects that lie significantly below the mass-metallicity relation in Figure \ref{mass_metal} have ratios of gas phase metallicity to stellar metallicity up to an order-of-magnitude lower, strongly suggesting that this material must have come from an external source. 

\subsection{Enrichment timescales}
\label{enrich}

   \begin{figure*}
 \begin{center}
 \includegraphics[height=6cm,angle=0,clip,trim=0cm 0cm 4.3cm 0.0cm]{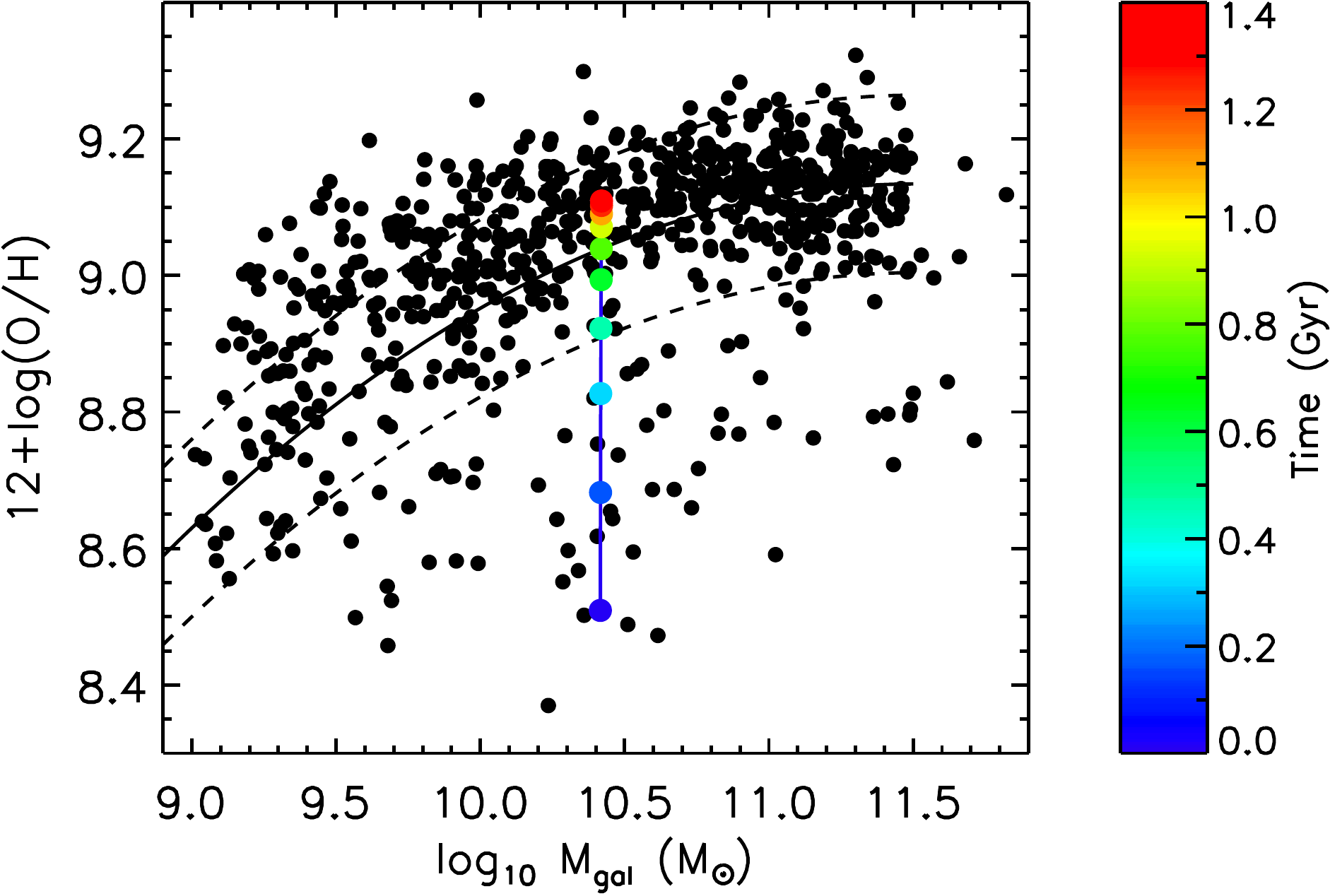}\hspace{0.25cm}
\includegraphics[height=6cm,angle=0,clip,trim=0cm 0cm 0cm 0.0cm]{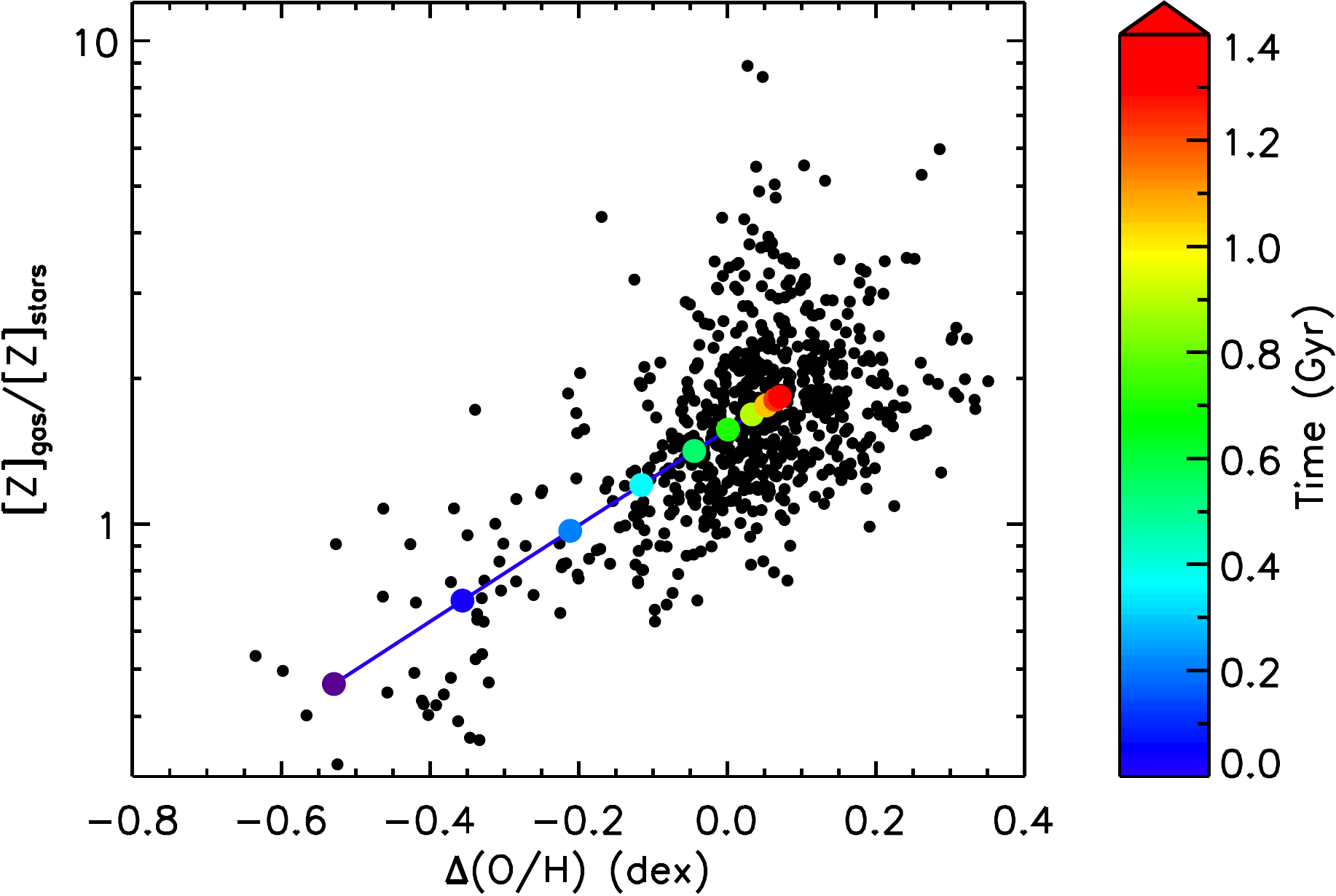}
\caption{Mass vs metallicity (left-hand panel) and residual metallicity vs enhancement of the gas-phase metallicity (right-hand panel), as in Figures \ref{mass_metal} and  \ref{mass_metal_wstellar}. The blue lines show the track followed by a typical galaxy from our observational sample (with a stellar mass of 2.6$\times10^{10}$ \msun and a stellar metallicity of 0.89 Z$_{\odot}$) as it undergoes a burst of star caused by 5$\times10^8$ \msun\ of cold gas which initially has a metallicity of 0.1 Z$_{\odot}$. Coloured points along this line, and the colour-bar on the far right, show how quickly our model galaxy moves in this space. The signature of metal-poor accretion is only visible for $\approx$400 Myr.  }
\label{deltastarmet_deltagasmet}
 \end{center}
   \vspace{-0.6cm}
 \end{figure*}

In the above sections we showed that $\approx$7.4\% of star-forming ETGs in our sample have gas-phase metallicities at least 2$\sigma$ below the \cite{2004ApJ...613..898T} relation, and that their gas is less enriched than their stellar bodies. However, this fraction of objects with visible low metallicity gas ($f_{\rm visible}$) is only a lower limit to the true importance of low-metallicity gas accretion in fuelling residual star formation in ETGs. As soon as low-metallicity gas starts forming stars in its host it will be quickly polluted by supernovae, and in a short time will be enriched enough to become indistinguishable from stellar mass loss material. Only by understanding the visibility timescale over which this happens ($t_{\rm visible}$) with respect to the total time a star formation episode takes ($t_{\rm total}$) can we attempt to correct for it, and thus reveal the true fraction ($f_{\rm true}$) of such accretion events. These variables are simply related as follows
\begin{equation}
\label{timescaleeqn}
f_{\rm true} =  \frac{t_{\rm total}}{t_{\rm visible}}f_{\rm visible}.
\end{equation}
In order to estimate the visibility timescale $t_{\rm visible}$ we used the chemical evolution model \textit{ChemEvol}\footnote{\url{https://github.com/zemogle/chemevol}} \citep{2003MNRAS.343..427M,2014MNRAS.441.1040R,2017MNRAS.471.1743D}. We simulated the chemical evolution of 5$\times10^8$ \msun\ of cold gas (the typical amount of H$_2$ found in gas rich ETGs; \citealt{2011MNRAS.414..940Y,2019MNRAS.486.1404D}) that initially has a metallicity of 0.1 Z$_{\odot}$, which falls into a typical galaxy from our observational sample (with a stellar mass of 2.6$\times10^{10}$ \msun and a stellar metallicity of 0.89 Z$_{\odot}$). The gas forms stars in a single episode, which we model using a gaussian star formation history with a standard deviation of 500 million years. This choice is somewhat arbitrary, but similar results are obtained if we were to instead assume an exponentially decaying star formation history. In our model this star formation does not drive a wind, and thus all the metals produced remain in the galaxy. As our galaxies are massive supernovae-driven winds struggle to remove significant amounts of gas, and thus including them would not significantly change the derived timescale.

 In Figure \ref{deltastarmet_deltagasmet} we show the track this model galaxy takes, over-plotted on our observational data from Figures \ref{mass_metal} and \ref{mass_metal_wstellar}. We show the evolution of the metallicity as a blue line, which begins at $t=0$, when the star formation rate in the system reaches 0.1 \msun\,yr$^{-1}$ (the lowest star formation rate found in our low metallicity ETGs). These tracks well match the distribution of the observed galaxies. The coloured points along this line show how quickly the galaxy moves through this space;  a galaxy takes $\approx$400 Myr to transition from having low gas-phase metallicity for its mass, to being fully consistent with the mass--metallicity relationship.  This defines the visibility timescale for our analysis, {which we note is very similar to the to the timescale for rejuvenated, blue ETGs to move back to the red sequence \citep[e.g.][]{2007MNRAS.382.1415S}.}   Assuming that ETGs form stars with the same depletion time as typical spirals ($t_{\rm total}=2$ Gyr; e.g. \citealt{2011ApJ...730L..13B}) we estimate using Equation \ref{timescaleeqn} that at least 37\% of gas rich ETGs have accreted their star forming gas from an external (low metallicity) source. If the true depletion time is longer in ETGs, as indicated by some analyses (e.g. \citealt{2011MNRAS.415...61S,2014MNRAS.444.3427D}), then this fraction would increase.

\section{Discussion and Conclusions}
\label{discuss}
\label{conclude}

In this letter we have shown that at fixed mass star-forming Galaxy Zoo elliptical galaxies show substantially different gas-phase metallicity distribution than star-forming spiral galaxies.
The `fundamental' mass -- metallicity relation (e.g. \citealt{2010MNRAS.408.2115M}) would predict that these low SFR ETGs would be more metal-rich than spiral galaxies. We do find a signature of this within our data, as elliptical galaxies have a higher median metallicity. However, there are significant numbers of elliptical galaxies with low gas-phase metallicity.  In our sample 7.4\% of star-forming ETGs have gas-phase metallicities at least 2 standard deviations below the \cite{2004ApJ...613..898T} mass -- metallicity relation, while only 1.7\% of spiral galaxies do. These systems typically have lower gas-phase metallicity than stellar metallicities, which strongly suggests that this material was accreted from an external source. We use a simple chemical evolution model to show that the visibility timescale for low-metallicity gas can be very short, and estimate that accretion events supply the gas in $\gtsimeq$37\% of star forming ETGs.

Various uncertainties may affect this result. For instance, we implement a simplistic toy model of a gas accretion event, and altering the assumed SFH, initial metallicity, IMF etc can alter the derived visibility timescale, and thus derived accretion fraction. Studying gas accretion events onto gas poor galaxies in cosmological simulations may provide additional insights into this process, and allow us to marginalise over these uncertainties.   Various effects were also not taken into account in the analysis above. For instance there is a known correlation between metallicity and large scale environment of galaxies \citep[e.g.][]{2008MNRAS.390..245C,2014MNRAS.438..262P}. By cross-matching with the group catalogue of \cite{2014A&A...566A...1T} we find, however, that our low-metallicity gas-rich ETGs are in the same environments as their high-metallicity counterparts, typically residing in the field and in galaxy groups. {This is consistent with the finding that both the integrated and resolved stellar population scaling relations in ETGs do not seem to depend strongly on environment (see e.g. \citealt{2010MNRAS.404.1775T,2017MNRAS.465..688G}). This also suggests the regenerative events we are probing with our star forming ETG sample here do not depend strongly on local environment, but may be suppressed in galaxy clusters (in agreement with the results of e.g. \citealt{2005ApJ...621..673T,2009MNRAS.396..818S,2011MNRAS.417..882D}).} 

A further uncertainty is that we are drawing conclusions about a population based upon a subset of its members. It is possible that ETGs with enough star formation for OB stars to dominate the ionisation of their warm ISM may not be representative of the underlying population. However, the study of \cite{2017MNRAS.466.2570B} suggests that any such effect is likely to cause us to underestimate the accreted fraction. In that work they use kinematic misalignments to constrain the origin of the gas, and find that star-formation dominated objects have a lower accreted fraction than objects where the ionisation is dominated by old stars/shocks.

Another explanation for our results could be that high metallicity stellar mass loss remains in the halo long enough to mix, and be efficiently diluted with significant quantities of lower metallicity material from further out in the circumgalactic medium around the galaxy. With the data we have in hand we cannot rule this possibility out. Resolved observations of the low gas-phase metallicity systems would determine if the gas also shows signatures of disturbance/misalignment, which could further constrain its origin (although misaligned hot halos may exist, which further complicates this analysis; \citealt{2015MNRAS.448.1271L}). Unfortunately to date none of these objects are included in the public data releases of the Mapping Nearby Galaxies at APO (MaNGA; \citealt{2015ApJ...798....7B}) survey. High-resolution simulations of the disc/halo interface could also be used to help determine how well mixed stellar mass loss becomes before it cools.

Despite these uncertainties our estimate that accretion dominates the gas supply in $\gtsimeq$37.5\% of ellipticals agrees well with those determined using complementary techniques, with different systematics. For instance, \cite{2011MNRAS.417..882D} used the kinematics of the cold and ionised ISM in a sample of 260 ETGs to show that at $\gtsimeq$42\% per cent of local fast-rotating early-type galaxies in the field have their gas supplied from external sources. \cite{2019MNRAS.483..458B} used the same technique in an even larger galaxy sample, and found evidence for external accretion in $\gtsimeq$45\% per cent of ETGs. \cite{2014MNRAS.437L..41K} use star formation rate distributions to estimate that $>$24\% of the star formation in ETGs has been driven by mergers/accretion events. 
Conversely, \cite{2019MNRAS.484..562G} used metallicity measurements of a small sample of 3 ETGs to argue that cooling from the hot halo was the more likely origin of the gas, even their two kinematically disturbed/misaligned systems. Given that only 7.4\% of ETGs show low gas-phase metallicities, and that star formation can enrich the gas quickly (see Section \ref{enrich}) it seems plausible that the difference between this study and our own is driven by the sample size. In all, the good agreement with past work gives us confidence that studying the gas-phase metallicity distribution allows us to obtain information about the origin of the gas in ETGs. 

Low gas-phase metallicity outliers from the mass metallicity relation are 4.4 times more common in the ETG population than in spiral galaxies. Given that (to first order) mergers and accretion events are equally likely to happen to each class of object it is interesting to consider why these rates are so different. We suggest that this likely arises because spiral galaxies are almost always ISM rich, while the majority of ellipticals are ISM poor. The mass of material required to dilute the metallicity of the existing reservoir in a galaxy is thus significantly larger in spiral galaxies than ETGs, requiring rarer high gas-fraction mergers.

Clearly to make further progress using this technique we will require large spectroscopic surveys from new instruments such as 4MOST (the 4-metre Multi-Object Spectroscopic Telescope; \citealt{2019Msngr.175....3D}) and MOONS (Multi Object Optical and Near-infrared Spectrograph; \citealt{2018SPIE10702E..1GT}), matched with high quality imaging to enable citizen science/machine learning morphology classifications for as many galaxies as possible. In addition, better methods for measuring gas-phase metallicities will allow us to make the most of existing observations. Pairing the above with resolved spectroscopy (in the optical, or of the cold gas components) of large galaxy samples will help us to definitively pin down the mechanisms supplying gas to ETGs, and their relative importance.

\vspace{0.25cm}
\noindent \textbf{Acknowledgments:}

\noindent We thank the referee (Daniel Thomas) for helpful comments which improved this manuscript. TAD acknowledges support from a Science and Technology Facilities Council Ernest Rutherford Fellowship. TAD thanks Freeke van de Voort for helpful comments on an early version of this letter, and Pieter De Vis for help with \textit{Chemevol}.
\\
\bsp
\bibliographystyle{mnras}
\bibliography{bib_gasz.bib}

\begin{thebibliography}{}
\makeatletter
\relax
\def\mn@urlcharsother{\let\do\@makeother \do\$\do\&\do\#\do\^\do\_\do\%\do\~}
\def\mn@doi{\begingroup\mn@urlcharsother \@ifnextchar [ {\mn@doi@}
  {\mn@doi@[]}}
\def\mn@doi@[#1]#2{\def\@tempa{#1}\ifx\@tempa\@empty \href
  {http://dx.doi.org/#2} {doi:#2}\else \href {http://dx.doi.org/#2} {#1}\fi
  \endgroup}
\def\mn@eprint#1#2{\mn@eprint@#1:#2::\@nil}
\def\mn@eprint@arXiv#1{\href {http://arxiv.org/abs/#1} {{\tt arXiv:#1}}}
\def\mn@eprint@dblp#1{\href {http://dblp.uni-trier.de/rec/bibtex/#1.xml}
  {dblp:#1}}
\def\mn@eprint@#1:#2:#3:#4\@nil{\def\@tempa {#1}\def\@tempb {#2}\def\@tempc
  {#3}\ifx \@tempc \@empty \let \@tempc \@tempb \let \@tempb \@tempa \fi \ifx
  \@tempb \@empty \def\@tempb {arXiv}\fi \@ifundefined
  {mn@eprint@\@tempb}{\@tempb:\@tempc}{\expandafter \expandafter \csname
  mn@eprint@\@tempb\endcsname \expandafter{\@tempc}}}

\bibitem[\protect\citeauthoryear{{Abazajian} et~al.,}{{Abazajian}
  et~al.}{2009}]{2009ApJS..182..543A}
{Abazajian} K.~N.,  et~al., 2009, \mn@doi [\apjs]
  {10.1088/0067-0049/182/2/543}, \href
  {https://ui.adsabs.harvard.edu/abs/2009ApJS..182..543A} {182, 543}

\bibitem[\protect\citeauthoryear{{Alatalo} et~al.,}{{Alatalo}
  et~al.}{2013}]{2013MNRAS.432.1796A}
{Alatalo} K.,  et~al., 2013, \mn@doi [\mnras] {10.1093/mnras/sts299}, \href
  {https://ui.adsabs.harvard.edu/abs/2013MNRAS.432.1796A} {432, 1796}

\bibitem[\protect\citeauthoryear{{Anders} \& {Grevesse}}{{Anders} \&
  {Grevesse}}{1989}]{1989GeCoA..53..197A}
{Anders} E.,  {Grevesse} N.,  1989, \mn@doi [\gca]
  {10.1016/0016-7037(89)90286-X}, \href
  {https://ui.adsabs.harvard.edu/abs/1989GeCoA..53..197A} {53, 197}

\bibitem[\protect\citeauthoryear{{Athey}, {Bregman}, {Bregman}, {Temi}  \&
  {Sauvage}}{{Athey} et~al.}{2002}]{2002ApJ...571..272A}
{Athey} A.,  {Bregman} J.,  {Bregman} J.,  {Temi} P.,   {Sauvage} M.,  2002,
  \mn@doi [\apj] {10.1086/339844}, \href
  {https://ui.adsabs.harvard.edu/abs/2002ApJ...571..272A} {571, 272}

\bibitem[\protect\citeauthoryear{{Bamford} et~al.,}{{Bamford}
  et~al.}{2009}]{2009MNRAS.393.1324B}
{Bamford} S.~P.,  et~al., 2009, \mn@doi [\mnras]
  {10.1111/j.1365-2966.2008.14252.x}, \href
  {https://ui.adsabs.harvard.edu/abs/2009MNRAS.393.1324B} {393, 1324}

\bibitem[\protect\citeauthoryear{{Belfiore} et~al.,}{{Belfiore}
  et~al.}{2016}]{2016MNRAS.461.3111B}
{Belfiore} F.,  et~al., 2016, \mn@doi [\mnras] {10.1093/mnras/stw1234}, \href
  {https://ui.adsabs.harvard.edu/abs/2016MNRAS.461.3111B} {461, 3111}

\bibitem[\protect\citeauthoryear{{Belfiore} et~al.,}{{Belfiore}
  et~al.}{2017}]{2017MNRAS.466.2570B}
{Belfiore} F.,  et~al., 2017, \mn@doi [\mnras] {10.1093/mnras/stw3211}, \href
  {https://ui.adsabs.harvard.edu/abs/2017MNRAS.466.2570B} {466, 2570}

\bibitem[\protect\citeauthoryear{{Bigiel} et~al.,}{{Bigiel}
  et~al.}{2011}]{2011ApJ...730L..13B}
{Bigiel} F.,  et~al., 2011, \mn@doi [\apjl] {10.1088/2041-8205/730/2/L13},
  \href {https://ui.adsabs.harvard.edu/abs/2011ApJ...730L..13B} {730, L13}

\bibitem[\protect\citeauthoryear{{Brinchmann}, {Charlot}, {White}, {Tremonti},
  {Kauffmann}, {Heckman}  \& {Brinkmann}}{{Brinchmann}
  et~al.}{2004}]{2004MNRAS.351.1151B}
{Brinchmann} J.,  {Charlot} S.,  {White} S.~D.~M.,  {Tremonti} C.,  {Kauffmann}
  G.,  {Heckman} T.,   {Brinkmann} J.,  2004, \mn@doi [\mnras]
  {10.1111/j.1365-2966.2004.07881.x}, \href
  {https://ui.adsabs.harvard.edu/abs/2004MNRAS.351.1151B} {351, 1151}

\bibitem[\protect\citeauthoryear{{Bryant} et~al.,}{{Bryant}
  et~al.}{2019}]{2019MNRAS.483..458B}
{Bryant} J.~J.,  et~al., 2019, \mn@doi [\mnras] {10.1093/mnras/sty3122}, \href
  {https://ui.adsabs.harvard.edu/abs/2019MNRAS.483..458B} {483, 458}

\bibitem[\protect\citeauthoryear{{Bundy} et~al.,}{{Bundy}
  et~al.}{2015}]{2015ApJ...798....7B}
{Bundy} K.,  et~al., 2015, \mn@doi [\apj] {10.1088/0004-637X/798/1/7}, \href
  {https://ui.adsabs.harvard.edu/abs/2015ApJ...798....7B} {798, 7}

\bibitem[\protect\citeauthoryear{{Cooper}, {Tremonti}, {Newman}  \&
  {Zabludoff}}{{Cooper} et~al.}{2008}]{2008MNRAS.390..245C}
{Cooper} M.~C.,  {Tremonti} C.~A.,  {Newman} J.~A.,   {Zabludoff} A.~I.,  2008,
  \mn@doi [\mnras] {10.1111/j.1365-2966.2008.13714.x}, \href
  {https://ui.adsabs.harvard.edu/abs/2008MNRAS.390..245C} {390, 245}

\bibitem[\protect\citeauthoryear{{Crocker}, {Bureau}, {Young}  \&
  {Combes}}{{Crocker} et~al.}{2011}]{2011MNRAS.410.1197C}
{Crocker} A.,  {Bureau} M.,  {Young} L.,   {Combes} F.,  2011, \mn@doi [\mnras]
  {10.1111/j.1365-2966.2010.17537.x}, \href
  {https://ui.adsabs.harvard.edu/abs/2011MNRAS.410.1197C} {410, 1197}

\bibitem[\protect\citeauthoryear{{Danforth} \& {Shull}}{{Danforth} \&
  {Shull}}{2008}]{2008ApJ...679..194D}
{Danforth} C.~W.,  {Shull} J.~M.,  2008, \mn@doi [\apj] {10.1086/587127}, \href
  {https://ui.adsabs.harvard.edu/abs/2008ApJ...679..194D} {679, 194}

\bibitem[\protect\citeauthoryear{{Davis} \& {Bureau}}{{Davis} \&
  {Bureau}}{2016}]{2016MNRAS.457..272D}
{Davis} T.~A.,  {Bureau} M.,  2016, \mn@doi [\mnras] {10.1093/mnras/stv2998},
  \href {https://ui.adsabs.harvard.edu/abs/2016MNRAS.457..272D} {457, 272}

\bibitem[\protect\citeauthoryear{{Davis} et~al.,}{{Davis}
  et~al.}{2011}]{2011MNRAS.417..882D}
{Davis} T.~A.,  et~al., 2011, \mn@doi [\mnras]
  {10.1111/j.1365-2966.2011.19355.x}, \href
  {https://ui.adsabs.harvard.edu/abs/2011MNRAS.417..882D} {417, 882}

\bibitem[\protect\citeauthoryear{{Davis} et~al.,}{{Davis}
  et~al.}{2013}]{2013MNRAS.429..534D}
{Davis} T.~A.,  et~al., 2013, \mn@doi [\mnras] {10.1093/mnras/sts353}, \href
  {https://ui.adsabs.harvard.edu/abs/2013MNRAS.429..534D} {429, 534}

\bibitem[\protect\citeauthoryear{{Davis} et~al.,}{{Davis}
  et~al.}{2014}]{2014MNRAS.444.3427D}
{Davis} T.~A.,  et~al., 2014, \mn@doi [MNRAS] {10.1093/mnras/stu570}, \href
  {http://adsabs.harvard.edu/abs/2014MNRAS.444.3427D} {444, 3427}

\bibitem[\protect\citeauthoryear{{Davis} et~al.,}{{Davis}
  et~al.}{2015}]{2015MNRAS.449.3503D}
{Davis} T.~A.,  et~al., 2015, \mn@doi [\mnras] {10.1093/mnras/stv597}, \href
  {https://ui.adsabs.harvard.edu/abs/2015MNRAS.449.3503D} {449, 3503}

\bibitem[\protect\citeauthoryear{{Davis}, {Greene}, {Ma}, {Blakeslee},
  {Dawson}, {Pandya}, {Veale}  \& {Zabel}}{{Davis}
  et~al.}{2019}]{2019MNRAS.486.1404D}
{Davis} T.~A.,  {Greene} J.~E.,  {Ma} C.-P.,  {Blakeslee} J.~P.,  {Dawson}
  J.~M.,  {Pandya} V.,  {Veale} M.,   {Zabel} N.,  2019, \mn@doi [\mnras]
  {10.1093/mnras/stz871}, \href
  {https://ui.adsabs.harvard.edu/abs/2019MNRAS.486.1404D} {486, 1404}

\bibitem[\protect\citeauthoryear{{De Vis} et~al.,}{{De Vis}
  et~al.}{2017}]{2017MNRAS.471.1743D}
{De Vis} P.,  et~al., 2017, \mn@doi [\mnras] {10.1093/mnras/stx981}, \href
  {https://ui.adsabs.harvard.edu/abs/2017MNRAS.471.1743D} {471, 1743}

\bibitem[\protect\citeauthoryear{{Faucher-Gigu{\`e}re}, {Kere{\v{s}}}  \&
  {Ma}}{{Faucher-Gigu{\`e}re} et~al.}{2011}]{2011MNRAS.417.2982F}
{Faucher-Gigu{\`e}re} C.-A.,  {Kere{\v{s}}} D.,   {Ma} C.-P.,  2011, \mn@doi
  [\mnras] {10.1111/j.1365-2966.2011.19457.x}, \href
  {https://ui.adsabs.harvard.edu/abs/2011MNRAS.417.2982F} {417, 2982}

\bibitem[\protect\citeauthoryear{{Goddard} et~al.,}{{Goddard}
  et~al.}{2017}]{2017MNRAS.465..688G}
{Goddard} D.,  et~al., 2017, \mn@doi [\mnras] {10.1093/mnras/stw2719}, \href
  {https://ui.adsabs.harvard.edu/abs/2017MNRAS.465..688G} {465, 688}

\bibitem[\protect\citeauthoryear{{Goudfrooij} \& {de Jong}}{{Goudfrooij} \& {de
  Jong}}{1995}]{1995A&A...298..784G}
{Goudfrooij} P.,  {de Jong} T.,  1995, \aap, \href
  {https://ui.adsabs.harvard.edu/abs/1995A&A...298..784G} {298, 784}

\bibitem[\protect\citeauthoryear{{Griffith}, {Martini}  \& {Conroy}}{{Griffith}
  et~al.}{2019}]{2019MNRAS.484..562G}
{Griffith} E.,  {Martini} P.,   {Conroy} C.,  2019, \mn@doi [\mnras]
  {10.1093/mnras/sty3405}, \href
  {https://ui.adsabs.harvard.edu/abs/2019MNRAS.484..562G} {484, 562}

\bibitem[\protect\citeauthoryear{{Guo}, {Zheng}, {Wang}  \& {Fu}}{{Guo}
  et~al.}{2015}]{2015ApJ...808L..49G}
{Guo} K.,  {Zheng} X.~Z.,  {Wang} T.,   {Fu} H.,  2015, \mn@doi [\apjl]
  {10.1088/2041-8205/808/2/L49}, \href
  {https://ui.adsabs.harvard.edu/abs/2015ApJ...808L..49G} {808, L49}

\bibitem[\protect\citeauthoryear{{Jungwiert}, {Combes}  \& {Palou{\v
  s}}}{{Jungwiert} et~al.}{2001}]{2001A&A...376...85J}
{Jungwiert} B.,  {Combes} F.,   {Palou{\v s}} J.,  2001, \mn@doi [\aap]
  {10.1051/0004-6361:20010966}, \href
  {https://ui.adsabs.harvard.edu/abs/2001A%26A...376...85J} {376, 85}

\bibitem[\protect\citeauthoryear{{Kauffmann} et~al.,}{{Kauffmann}
  et~al.}{2003a}]{2003MNRAS.341...33K}
{Kauffmann} G.,  et~al., 2003a, \mn@doi [\mnras]
  {10.1046/j.1365-8711.2003.06291.x}, \href
  {https://ui.adsabs.harvard.edu/abs/2003MNRAS.341...33K} {341, 33}

\bibitem[\protect\citeauthoryear{{Kauffmann} et~al.,}{{Kauffmann}
  et~al.}{2003b}]{2003MNRAS.346.1055K}
{Kauffmann} G.,  et~al., 2003b, \mn@doi [\mnras]
  {10.1111/j.1365-2966.2003.07154.x}, \href
  {https://ui.adsabs.harvard.edu/abs/2003MNRAS.346.1055K} {346, 1055}

\bibitem[\protect\citeauthoryear{{Kaviraj}}{{Kaviraj}}{2014}]{2014MNRAS.437L..41K}
{Kaviraj} S.,  2014, \mn@doi [\mnras] {10.1093/mnrasl/slt136}, \href
  {https://ui.adsabs.harvard.edu/abs/2014MNRAS.437L..41K} {437, L41}

\bibitem[\protect\citeauthoryear{{Kaviraj} et~al.,}{{Kaviraj}
  et~al.}{2007}]{2007ApJS..173..619K}
{Kaviraj} S.,  et~al., 2007, \mn@doi [\apjs] {10.1086/516633}, \href
  {https://ui.adsabs.harvard.edu/abs/2007ApJS..173..619K} {173, 619}

\bibitem[\protect\citeauthoryear{{Kewley} \& {Ellison}}{{Kewley} \&
  {Ellison}}{2008}]{2008ApJ...681.1183K}
{Kewley} L.~J.,  {Ellison} S.~L.,  2008, \mn@doi [\apj] {10.1086/587500}, \href
  {https://ui.adsabs.harvard.edu/abs/2008ApJ...681.1183K} {681, 1183}

\bibitem[\protect\citeauthoryear{{Lagos}, {Davis}, {Lacey}, {Zwaan}, {Baugh},
  {Gonzalez-Perez}  \& {Padilla}}{{Lagos} et~al.}{2014}]{2014MNRAS.443.1002L}
{Lagos} C.~d.~P.,  {Davis} T.~A.,  {Lacey} C.~G.,  {Zwaan} M.~A.,  {Baugh}
  C.~M.,  {Gonzalez-Perez} V.,   {Padilla} N.~D.,  2014, \mn@doi [\mnras]
  {10.1093/mnras/stu1209}, \href
  {https://ui.adsabs.harvard.edu/abs/2014MNRAS.443.1002L} {443, 1002}

\bibitem[\protect\citeauthoryear{{Lagos}, {Padilla}, {Davis}, {Lacey}, {Baugh},
  {Gonzalez-Perez}, {Zwaan}  \& {Contreras}}{{Lagos}
  et~al.}{2015}]{2015MNRAS.448.1271L}
{Lagos} C.~d.~P.,  {Padilla} N.,  {Davis} T.~A.,  {Lacey} C.,  {Baugh} C.,
  {Gonzalez-Perez} V.,  {Zwaan} M.,   {Contreras} S.,  2015, \mn@doi [\mnras]
  {10.1093/mnras/stu2763}, \href
  {https://ui.adsabs.harvard.edu/abs/2015MNRAS.448.1271L} {448, 1271}

\bibitem[\protect\citeauthoryear{{Li} et~al.,}{{Li}
  et~al.}{2018}]{2018ApJ...866...70L}
{Li} Y.-P.,  et~al., 2018, \mn@doi [\apj] {10.3847/1538-4357/aade8b}, \href
  {https://ui.adsabs.harvard.edu/abs/2018ApJ...866...70L} {866, 70}

\bibitem[\protect\citeauthoryear{{Lianou}, {Barmby}, {Mosenkov}, {Lehnert}  \&
  {Karczewski}}{{Lianou} et~al.}{2019}]{2019arXiv190602712L}
{Lianou} S.,  {Barmby} P.,  {Mosenkov} A.,  {Lehnert} M.,   {Karczewski} O.,
  2019, arXiv:1906.02712, \href
  {https://ui.adsabs.harvard.edu/abs/2019arXiv190602712L} {}

\bibitem[\protect\citeauthoryear{{Lintott} et~al.,}{{Lintott}
  et~al.}{2008}]{2008MNRAS.389.1179L}
{Lintott} C.~J.,  et~al., 2008, \mn@doi [\mnras]
  {10.1111/j.1365-2966.2008.13689.x}, \href
  {https://ui.adsabs.harvard.edu/abs/2008MNRAS.389.1179L} {389, 1179}

\bibitem[\protect\citeauthoryear{{Lintott} et~al.,}{{Lintott}
  et~al.}{2011}]{2011MNRAS.410..166L}
{Lintott} C.,  et~al., 2011, \mn@doi [\mnras]
  {10.1111/j.1365-2966.2010.17432.x}, \href
  {https://ui.adsabs.harvard.edu/abs/2011MNRAS.410..166L} {410, 166}

\bibitem[\protect\citeauthoryear{{Mannucci}, {Cresci}, {Maiolino}, {Marconi}
  \& {Gnerucci}}{{Mannucci} et~al.}{2010}]{2010MNRAS.408.2115M}
{Mannucci} F.,  {Cresci} G.,  {Maiolino} R.,  {Marconi} A.,   {Gnerucci} A.,
  2010, \mn@doi [\mnras] {10.1111/j.1365-2966.2010.17291.x}, \href
  {https://ui.adsabs.harvard.edu/abs/2010MNRAS.408.2115M} {408, 2115}

\bibitem[\protect\citeauthoryear{{Maraston} \& {Str{\"o}mb{\"a}ck}}{{Maraston}
  \& {Str{\"o}mb{\"a}ck}}{2011}]{2011MNRAS.418.2785M}
{Maraston} C.,  {Str{\"o}mb{\"a}ck} G.,  2011, \mn@doi [\mnras]
  {10.1111/j.1365-2966.2011.19738.x}, \href
  {https://ui.adsabs.harvard.edu/abs/2011MNRAS.418.2785M} {418, 2785}

\bibitem[\protect\citeauthoryear{{Martig} et~al.,}{{Martig}
  et~al.}{2013}]{2013MNRAS.432.1914M}
{Martig} M.,  et~al., 2013, \mn@doi [\mnras] {10.1093/mnras/sts594}, \href
  {https://ui.adsabs.harvard.edu/abs/2013MNRAS.432.1914M} {432, 1914}

\bibitem[\protect\citeauthoryear{{Morgan} \& {Edmunds}}{{Morgan} \&
  {Edmunds}}{2003}]{2003MNRAS.343..427M}
{Morgan} H.~L.,  {Edmunds} M.~G.,  2003, \mn@doi [\mnras]
  {10.1046/j.1365-8711.2003.06681.x}, \href
  {https://ui.adsabs.harvard.edu/abs/2003MNRAS.343..427M} {343, 427}

\bibitem[\protect\citeauthoryear{{Nyland} et~al.,}{{Nyland}
  et~al.}{2017}]{2017MNRAS.464.1029N}
{Nyland} K.,  et~al., 2017, \mn@doi [\mnras] {10.1093/mnras/stw2385}, \href
  {https://ui.adsabs.harvard.edu/abs/2017MNRAS.464.1029N} {464, 1029}

\bibitem[\protect\citeauthoryear{{Peng} \& {Maiolino}}{{Peng} \&
  {Maiolino}}{2014}]{2014MNRAS.438..262P}
{Peng} Y.-j.,  {Maiolino} R.,  2014, \mn@doi [\mnras] {10.1093/mnras/stt2175},
  \href {https://ui.adsabs.harvard.edu/abs/2014MNRAS.438..262P} {438, 262}

\bibitem[\protect\citeauthoryear{{Pulido} et~al.,}{{Pulido}
  et~al.}{2018}]{2018ApJ...853..177P}
{Pulido} F.~A.,  et~al., 2018, \mn@doi [\apj] {10.3847/1538-4357/aaa54b}, \href
  {https://ui.adsabs.harvard.edu/abs/2018ApJ...853..177P} {853, 177}

\bibitem[\protect\citeauthoryear{{Rowlands}, {Gomez}, {Dunne},
  {Arag{\'o}n-Salamanca}, {Dye}, {Maddox}, {da Cunha}  \& {van der
  Werf}}{{Rowlands} et~al.}{2014}]{2014MNRAS.441.1040R}
{Rowlands} K.,  {Gomez} H.~L.,  {Dunne} L.,  {Arag{\'o}n-Salamanca} A.,  {Dye}
  S.,  {Maddox} S.,  {da Cunha} E.,   {van der Werf} P.,  2014, \mn@doi
  [\mnras] {10.1093/mnras/stu605}, \href
  {https://ui.adsabs.harvard.edu/abs/2014MNRAS.441.1040R} {441, 1040}

\bibitem[\protect\citeauthoryear{{Saintonge} et~al.,}{{Saintonge}
  et~al.}{2011}]{2011MNRAS.415...61S}
{Saintonge} A.,  et~al., 2011, \mn@doi [\mnras]
  {10.1111/j.1365-2966.2011.18823.x}, \href
  {https://ui.adsabs.harvard.edu/abs/2011MNRAS.415...61S} {415, 61}

\bibitem[\protect\citeauthoryear{{Salpeter}}{{Salpeter}}{1955}]{1955ApJ...121..161S}
{Salpeter} E.~E.,  1955, \mn@doi [\apj] {10.1086/145971}, \href
  {https://ui.adsabs.harvard.edu/abs/1955ApJ...121..161S} {121, 161}

\bibitem[\protect\citeauthoryear{{Sarzi} et~al.,}{{Sarzi}
  et~al.}{2006}]{2006MNRAS.366.1151S}
{Sarzi} M.,  et~al., 2006, \mn@doi [\mnras] {10.1111/j.1365-2966.2005.09839.x},
  \href {https://ui.adsabs.harvard.edu/abs/2006MNRAS.366.1151S} {366, 1151}

\bibitem[\protect\citeauthoryear{{Sarzi} et~al.,}{{Sarzi}
  et~al.}{2010}]{2010MNRAS.402.2187S}
{Sarzi} M.,  et~al., 2010, \mn@doi [\mnras] {10.1111/j.1365-2966.2009.16039.x},
  \href {https://ui.adsabs.harvard.edu/abs/2010MNRAS.402.2187S} {402, 2187}

\bibitem[\protect\citeauthoryear{{Schawinski}, {Thomas}, {Sarzi}, {Maraston},
  {Kaviraj}, {Joo}, {Yi}  \& {Silk}}{{Schawinski}
  et~al.}{2007}]{2007MNRAS.382.1415S}
{Schawinski} K.,  {Thomas} D.,  {Sarzi} M.,  {Maraston} C.,  {Kaviraj} S.,
  {Joo} S.-J.,  {Yi} S.~K.,   {Silk} J.,  2007, \mn@doi [\mnras]
  {10.1111/j.1365-2966.2007.12487.x}, \href
  {https://ui.adsabs.harvard.edu/abs/2007MNRAS.382.1415S} {382, 1415}

\bibitem[\protect\citeauthoryear{{Schawinski} et~al.,}{{Schawinski}
  et~al.}{2009}]{2009MNRAS.396..818S}
{Schawinski} K.,  et~al., 2009, \mn@doi [\mnras]
  {10.1111/j.1365-2966.2009.14793.x}, \href
  {https://ui.adsabs.harvard.edu/abs/2009MNRAS.396..818S} {396, 818}

\bibitem[\protect\citeauthoryear{{Smith} et~al.,}{{Smith}
  et~al.}{2012}]{2012ApJ...748..123S}
{Smith} M.~W.~L.,  et~al., 2012, \mn@doi [\apj] {10.1088/0004-637X/748/2/123},
  \href {https://ui.adsabs.harvard.edu/abs/2012ApJ...748..123S} {748, 123}

\bibitem[\protect\citeauthoryear{{Stasinska}}{{Stasinska}}{2019}]{2019arXiv190604520S}
{Stasinska} G.,  2019, in Chemical Abundances in Gaseous Nebulae: Open problems
  in Nebular astrophysics. Cardaci M., Hagele G., Perez-Montero E., eds,
  Asociaci\'on Argentina de Astronom\'ia Workshop Series.
 (\mn@eprint {} {1906.04520})

\bibitem[\protect\citeauthoryear{{Taylor} et~al.,}{{Taylor}
  et~al.}{2018}]{2018SPIE10702E..1GT}
{Taylor} W.,  et~al., 2018, in Ground-based and Airborne Instrumentation for
  Astronomy VII, Society of Photo-Optical Instrumentation Engineers (SPIE)
  Conference Series. p. 107021G, \mn@doi{10.1117/12.2313403}

\bibitem[\protect\citeauthoryear{{Temi}, {Brighenti}  \& {Mathews}}{{Temi}
  et~al.}{2009}]{2009ApJ...695....1T}
{Temi} P.,  {Brighenti} F.,   {Mathews} W.~G.,  2009, \mn@doi [\apj]
  {10.1088/0004-637X/695/1/1}, \href
  {https://ui.adsabs.harvard.edu/abs/2009ApJ...695....1T} {695, 1}

\bibitem[\protect\citeauthoryear{{Tempel} et~al.,}{{Tempel}
  et~al.}{2014}]{2014A&A...566A...1T}
{Tempel} E.,  et~al., 2014, \mn@doi [\aap] {10.1051/0004-6361/201423585}, \href
  {https://ui.adsabs.harvard.edu/abs/2014A&A...566A...1T} {566, A1}

\bibitem[\protect\citeauthoryear{{Thomas}, {Maraston}, {Bender}  \& {Mendes de
  Oliveira}}{{Thomas} et~al.}{2005}]{2005ApJ...621..673T}
{Thomas} D.,  {Maraston} C.,  {Bender} R.,   {Mendes de Oliveira} C.,  2005,
  \mn@doi [\apj] {10.1086/426932}, \href
  {https://ui.adsabs.harvard.edu/abs/2005ApJ...621..673T} {621, 673}

\bibitem[\protect\citeauthoryear{{Thomas}, {Maraston}, {Schawinski}, {Sarzi}
  \& {Silk}}{{Thomas} et~al.}{2010}]{2010MNRAS.404.1775T}
{Thomas} D.,  {Maraston} C.,  {Schawinski} K.,  {Sarzi} M.,   {Silk} J.,  2010,
  \mn@doi [\mnras] {10.1111/j.1365-2966.2010.16427.x}, \href
  {https://ui.adsabs.harvard.edu/abs/2010MNRAS.404.1775T} {404, 1775}

\bibitem[\protect\citeauthoryear{{Tremonti} et~al.,}{{Tremonti}
  et~al.}{2004}]{2004ApJ...613..898T}
{Tremonti} C.~A.,  et~al., 2004, \mn@doi [\apj] {10.1086/423264}, \href
  {https://ui.adsabs.harvard.edu/abs/2004ApJ...613..898T} {613, 898}

\bibitem[\protect\citeauthoryear{{Wilkinson}, {Maraston}, {Goddard}, {Thomas}
  \& {Parikh}}{{Wilkinson} et~al.}{2017}]{2017MNRAS.472.4297W}
{Wilkinson} D.~M.,  {Maraston} C.,  {Goddard} D.,  {Thomas} D.,   {Parikh} T.,
  2017, \mn@doi [\mnras] {10.1093/mnras/stx2215}, \href
  {https://ui.adsabs.harvard.edu/abs/2017MNRAS.472.4297W} {472, 4297}

\bibitem[\protect\citeauthoryear{{Young} et~al.,}{{Young}
  et~al.}{2011}]{2011MNRAS.414..940Y}
{Young} L.~M.,  et~al., 2011, \mn@doi [\mnras]
  {10.1111/j.1365-2966.2011.18561.x}, \href
  {https://ui.adsabs.harvard.edu/abs/2011MNRAS.414..940Y} {414, 940}

\bibitem[\protect\citeauthoryear{{Young} et~al.,}{{Young}
  et~al.}{2014}]{2014MNRAS.444.3408Y}
{Young} L.~M.,  et~al., 2014, \mn@doi [MNRAS] {10.1093/mnras/stt2474}, \href
  {http://adsabs.harvard.edu/abs/2014MNRAS.444.3408Y} {444, 3408}

\bibitem[\protect\citeauthoryear{{de Jong} et~al.,}{{de Jong}
  et~al.}{2019}]{2019Msngr.175....3D}
{de Jong} R.~S.,  et~al., 2019, \mn@doi [The Messenger]
  {10.18727/0722-6691/5117}, \href
  {https://ui.adsabs.harvard.edu/abs/2019Msngr.175....3D} {175, 3}

\bibitem[\protect\citeauthoryear{{van de Voort}, {Schaye}, {Booth}, {Haas}  \&
  {Dalla Vecchia}}{{van de Voort} et~al.}{2011}]{2011MNRAS.414.2458V}
{van de Voort} F.,  {Schaye} J.,  {Booth} C.~M.,  {Haas} M.~R.,   {Dalla
  Vecchia} C.,  2011, \mn@doi [\mnras] {10.1111/j.1365-2966.2011.18565.x},
  \href {https://ui.adsabs.harvard.edu/abs/2011MNRAS.414.2458V} {414, 2458}

\bibitem[\protect\citeauthoryear{{van de Voort} et~al.,}{{van de Voort}
  et~al.}{2018}]{2018MNRAS.476..122V}
{van de Voort} F.,  et~al., 2018, \mn@doi [\mnras] {10.1093/mnras/sty228},
  \href {https://ui.adsabs.harvard.edu/abs/2018MNRAS.476..122V} {476, 122}

\makeatother
\end{thebibliography}
\bibdata{bib_gasz.bib}
\bibstyle{mnras}

\label{lastpage}
\end{document}